\documentclass[12pt]{article}
\usepackage{amssymb,amsmath}
\usepackage[noblocks]{authblk}
\usepackage[top=0.75in, bottom=0.75in, left=0.75in, right=0.75in, dvips]{geometry}
\usepackage{caption}
\begin{document}
\textwidth 10.0in 
\textheight 9.0in 
\topmargin -0.60in
\title{The Renormalization Group and the Effective\\
Potential in the Wess-Zumino model}
\author[1,2]{D.G.C. McKeon}
\affil[1] {Department of Applied Mathematics, The
University of Western Ontario, London, ON N6A 5B7, Canada} 
\affil[2] {Department of Mathematics and
Computer Science, Algoma University, \newline Sault Ste. Marie, ON P6A
2G4, Canada}
\maketitle          

\maketitle      
\noindent
email: dgmckeo2@uwo.ca

\begin{abstract}

We consider the effective potential $V$ in the massless Wess-Zumino model.  By using the renormalization group equation, we show that the explicit and implicit dependence of $V$ on the renormalization mass scale $\mu$ cancels. If $V$ has an extremum at some non-vanishing value of the background field, then it follows that $V$ is ``flat'', independent of the background field.  This consistent with the general requirement that $V$ be convex.  The consequences for supersymmetric gauge theories are briefly considered.
\end{abstract}

\section{Introduction}

Spontaneous symmetry breaking [1,2] and the Higgs mechanism for mass generation [3,4,5] have played a central role in the Standard Model [6,7], cosmology [8] and supersymmetry [9].  Originally this was discussed using the classical effective potential $V$, but quantum contributions to $V$ were soon considered [10,11,12,13]. The need to renormalize to eliminate divergences when computing these quantum contributions resulted in introduction of a renormalization mass scale $\mu$ whose value is arbitrary; the renormalization group (RG) equation follows from the requirement that a physical quantity should be independent of $\mu$ [14,15,16].  It has been shown that by using the RG equation to sum the explicit dependence of a perturbative calculation on $\mu$, the explicit dependence on $\mu$ cancels against the implicit dependence on $\mu$ through the running parameters that characterize a theory [17,18,19].  In a number of models, when the RG equation is combined with the requirement that $V$ have an extremum at a non-zero value of the constant external field, it follows that $V$ is independent of this external field; $V$ is ``flat'' [20,21,22,23,24] and so the value of the external field is not fixed by minimizing $V$.  In this paper we show how this happens in the supersymmetric Wess-Zumino (WZ) model. (For useful reviews of supersymmetry, see for example refs. [25,26].)

\section{The Effective Potential in the Wess-Zumino Model}

The $n$-loop corrections to $V$ in general involve up to $n$ powered of $\ln \mu$.  The RG equation can be used to sum these logarithms; one loop contributions to the RG functions are needed to sum leading-log (LL) effects etc. [27,28,29]. Furthermore, all logarithmic contributions to $V$ can be summed leaving $V$ expressed in terms of its log-independent piece with all dependence on $\mu$ cancelling [24].

We will examine how the RG equation can be applied to $V$ in the WZ model.  The perturbative contributions to $V$ have been computed to one [30,31,32,33,34,35], two [36,37,38,39] and three [40] loop order, and RG summation of $LL$ contributions also examined [41].

In the WZ model, there is a chiral superfield $\Phi$ which involves a complex scalar field $\phi(x)$, a Majorana spinor field $\psi(x)$ and a complex auxiliary scalar field $f(x)$.  The interactions in this model are characterized by a ``superpotential''
\begin{equation}
P(\Phi) = m^2\Phi^2 + \lambda\Phi^3.
\end{equation}
We will restrict our attention to the massless case $(m^2 = 0)$ and take the background field $a$ to be constant and real.  Since $P(\Phi)$ is not altered by quantum corrections [42,43], $\lambda a^3$ is not renormalized and so if $\mu$ is the renormalization scale then 
\begin{equation}
\mu^2 \frac{d}{d\mu^2}(\lambda a^3) = 0.
\end{equation}
When one uses mass independent renormalization [44,45]
\begin{subequations}
\begin{align}
\mu^2 \frac{d\lambda}{d\mu^2} &= \beta(\lambda)\\
\mu^2 \frac{da^2}{d\mu^2} &= a^2\gamma(\lambda),
\end{align}
\end{subequations}
and eq. (2) implies that
\begin{equation}
\gamma(\lambda) = -\frac{2}{3\lambda}\beta(\lambda).
\end{equation}

The general form of $V$ is 
\begin{equation}
V(\lambda, a^2, \mu^2) = \sum_{n=0}^\infty \sum_{m=0}^n a^4 T_{n,m} \lambda^{n+1} \ln^m\left(\frac{\lambda a^2}{\mu^2}\right)
\end{equation}
where $T_{n,m}$ is computed by doing an $n$-loop calculation.  A general theorem states that if supersymmetry (SUSY) is unbroken, then at a value $a_0$ that minimizes $V$ [46],
\begin{equation}
V(\lambda, a_0^2, \mu^2) = 0.
\end{equation}
Furthermore, if $V$ is unbroken at a classical level when $a = a_0$, then this is also true when quantum effects are included [9,48,49,50].

We now define [20,21,22,23,24]
\begin{equation}
A_n(\lambda) = \sum_{k=0}^\infty T_{k+n,n} \lambda^{k+n+1} \quad (n = 0,1,\ldots)
\end{equation}
so that by eqs. (5,7)
\begin{equation}
V(\lambda, a, \mu^2) = \sum_{n=0}^\infty a^4 A_n(\lambda) \ln^n \left(\frac{\lambda a^2}{\mu^2}\right).
\end{equation}
Since $\mu^2$ is unphysical, we must have
\begin{align}
\mu^2 \frac{dV(\lambda, a^2, \mu^2)}{d\mu^2} &= 0\nonumber \\
&= \left( \mu^2 \frac{\partial}{\partial\mu^2} + \beta(\lambda) \frac{\partial}{\partial\lambda} + a^2 \gamma(\lambda) \frac{\partial}{\partial a^2}\right)V
\end{align}
which by eq. (4) becomes
\begin{equation}
\left[\mu^2 \frac{\partial}{\partial \mu^2} + \beta(\lambda) \left( \frac{\partial}{\partial\lambda} - \frac{2a^2}{3\lambda}\frac{\partial}{\partial a^2}\right)\right] V(\lambda, a^2, \mu^2) = 0.
\end{equation}
Substitution of eq. (8) into eq. (10) leads to 
\begin{equation}
A_n (\lambda) = \frac{1}{n}\frac{\beta(\lambda)}{1-\frac{\beta(\lambda)}{3\lambda}} \left( \frac{\partial}{\partial\lambda} - \frac{4}{3\lambda}\right) A_{n-1}(\lambda),
\end{equation}
so if
\begin{subequations}
\begin{align}
A_n (\lambda) &= \left[ \exp \int_{\lambda_0}^\lambda \left(\frac{4}{3x}\right)\right]B_n(\lambda)\\
\intertext{and}
\eta &= \int^\lambda_{\lambda_1} dx \frac{1-\frac{\beta(x)}{3x}}{\beta(x)}
\end{align}
\end{subequations}
($\lambda_i$ is a boundary value) then eq. (11) becomes
\begin{align}
B_n(\lambda(\eta)) &= \frac{1}{n} \frac{d}{d\eta} B_{n-1} (\lambda(\eta))\nonumber \\
&= \frac{1}{n!}\frac{d^n}{d\eta^n} B_0(\lambda(\eta)).
\end{align}
Together, eqs. (12a,13) result in eq. (8) becoming
\begin{align}
V(\lambda, a, \mu^2) = \sum_{n=0}^\infty & a^4 \frac{L^n}{n!} \left[\exp \int_{\lambda_0}^\lambda dx \left(\frac{4}{3x}\right)\right]
 \frac{d^n}{d\eta^n}\left[\exp - \int_{\lambda_0}^{\lambda(\eta)} dx \left(\frac{4}{3x}\right)\right]A_0(\lambda(\eta))\nonumber \\
&\hspace{-1cm} = a^4\left[ \exp - \int_\lambda^{\lambda(\eta+L)} dx \left(\frac{4}{3x}\right)\right] A_0(\lambda(\eta + L))
\end{align}
where $L \equiv \ln \left(\frac{\lambda a^2}{\mu^2}\right)$.

The solutions to eqs. (3a,3b) are written as 
\begin{subequations}
\begin{align}
\ln \left(\frac{\mu^2}{\Lambda^2}\right) &= \int_{\lambda_2}^{\lambda\left(\ln\frac{\mu^2}{\Lambda^2}\right)} \frac{dx}{\beta(x)}\\
a^2 &= \Phi^2 \left[\exp -   \int_{\lambda_3}^{\lambda\left(\ln\frac{\mu^2}{\Lambda^2}\right)} \frac{2dx}{3x}\right]
\end{align}
\end{subequations}
were $\Lambda^2$ and $\Phi^2$ are the values of $\mu^2$ and $a^2$ when $\lambda = \lambda_2$ and $\lambda = \lambda_3$ respectively. It follows from eq. (12b) that
\begin{equation}
\eta + L = \int_{\lambda_1}^\lambda \frac{dx}{\beta(x)}- \frac{1}{3}\int_{\lambda_1}^\lambda \frac{dx}{x} + \ln \lambda + \ln \left(\frac{a^2}{\mu^2}\right)
\end{equation}
which by eqs. (15a,b) becomes
\begin{align}
= \ln \left(\frac{\mu^2}{\Lambda^2}\right) & + \int_{\lambda_1}^{\lambda_2} \frac{dx}{\beta(x)} - \frac{1}{3}(\ln \lambda - \ln \lambda_1) + \ln\lambda\nonumber \\
&+ \ln \left(\frac{\Phi^2}{\mu^2}\right) - \frac{2}{3}(\ln \lambda - \ln \lambda_3)  \nonumber \\
& \hspace{-2.1cm} \equiv \ln  \left(\frac{\Phi^2}{\Lambda^2}\right) + K
\end{align}
where $K$ is a constant that can be absorbed into $\Phi^2/\Lambda^2$.  We see by eq. (17) that eq. (14) becomes
\begin{equation}
V = a^4 \left[ \exp - \int_\lambda^{\lambda\left(\ln \frac{\Phi^2}{\Lambda^2}\right)}
dx \left( \frac{4}{3x}\right) \right] A_0 \left( \lambda\left( \ln \frac{\Phi^2}{\Lambda^2}\right)\right).
\end{equation}
By eq. (15b) we can write eq. (18)
\begin{equation}
V = \Phi^4 \left[ \exp - \int_{\lambda_3}^{\lambda\left(\ln \frac{\Phi^2}{\Lambda^2}\right)}
dx \left( \frac{4}{3x}\right) \right] A_0 \left( \lambda\left( \ln \frac{\Phi^2}{\Lambda^2}\right)\right).
\end{equation}
In this expression for $V$, there is no ambiguity residing in the renormalization scale parameter $\mu^2$ as the explicit dependence of $V$ on $\mu^2$ (through $L$) and the implicit dependence on $\mu^2$ (through $\lambda$) has cancelled in eq. (17).

Having used the RG equation of eq. (9) to obtain eq. (19), we can now impose a second condition on $V$ to obtain the function $A_0$, the log independent terms in the expansion of eq. (8). If $V$ has an extremum at $\Phi^2 = \Phi^2_0$, then
\begin{equation}
\Phi^2 \frac{d}{d\Phi^2}\Bigg|_{\Phi_0^2} V = 0.
\end{equation}
Using eqs. (3a,19), eq. (20) becomes
\begin{equation}
\Phi_0^4  \left[ \exp - \int_{\lambda_3}^{\lambda\left(\ln \frac{\Phi^2_0}{\Lambda^2}\right)}
 \left( \frac{4}{3x}\right) \right] \left[2 A_0 (\lambda) - \frac{4}{3\lambda}\beta (\lambda) A_0 (\lambda) + \beta A_0^\prime (\lambda) \right] = 0.
\end{equation}
Eq. (21) is satisfied if either $\Phi_0^2 = 0$, or if
\begin{equation}
A_0^\prime + \left( \frac{2}{\beta} - \frac{4}{3\lambda}\right)A_0 = 0
\end{equation}
in which case
\begin{equation}
A_0 = \vert{\!\!\!C} \exp - \int_{\lambda_3}^\lambda  dx \left ( \frac{2}{\beta (x)} - \frac{4}{3x}
\right),
\end{equation}
where $\vert{\!\!\!C}$  is a constant of integration. Now substituting eq. (23) into eq. (19) we obtain
\begin{align}
V= \Phi^4 \Bigg[ \exp & - \int_{\lambda_3}^{\lambda \left(\ln \frac{\Phi^2}{\Lambda^2}\right)} dx  \left( \frac{4}{3x} \right) \Bigg] \nonumber \\
&\times \;\vert{\!\!\!C} \left[ \exp - \int_{\lambda_3}^\lambda dx \left(\frac{2}{\beta (x)} - \frac{4}{3x}\right)\right]\nonumber
\end{align}
which, when we use eq. (15a), becomes
\begin{align}
V &= \vert{\!\!\!C} \Phi^4 \left( \exp \ln \left(\frac{\Phi^2}{\Lambda^2}\right)^{-2}\right)\nonumber \\
 &= \vert{\!\!\!C} \Lambda^4.
\end{align}
We thus see that if $\Phi^2_0 \neq 0$, $V$ is independent of $\Phi^2$; it is flat.  This is consistent with the general requirement that $V$ must be convex [51,52]. It does not preclude spontaneous symmetry breaking, it just means that the vacuum expectation value of $V$ cannot be determined by minimizing $V$.

\section{Discussion}

We have shown that in the massless WZ model, either there is no spontaneous symmetry breaking (ie, $\Phi_0^2 = 0$) or that the effective potential is independent of the constant background field $\Phi^2$. The general theorem that if there is no spontaneous breakdown of SUSY at the classical level, then it remains unbroken by quantum effects means that since the classical potential in the massless WZ model does not break SUSY, then in eq. (24) we must take 
\begin{equation}
\vert{\!\!\!C} = 0
\end{equation}
in order for the full effective potential to leave SUSY unbroken.

The derivation of $V$ given in eq. (24) should be applicable to any massless supersymmetric theory which involves only one coupling.  In particular, both $N = 1$ supersymmetric gauge theories with no masses and $N = 2$ supersymmetric gauge theories are characterized by a single coupling, and so in both of these models, the effective potential $V$ is flat. 

The effective potential $V$ that we have been discussing is found by summing all renormalized one-particle irreducible diagrams with external scalar particles that have vanishing momentum.  This is not the same as the Wilsonian effective potential which is found by integrating out all high energy modes of the quantum fields. These two effective potentials differ in massless theories on account of infrared contributions.  Seiberg and Witten have used duality to analyze the Wilsonian low energy effective action of an $N = 2$ supersymmetric $SU(2)$ gauge theory; their analysis is quite distinct from what has been done in this paper [53].

\section*{Acknowledgements}
R. Macleod was quite helpful in this work.

\end{document}